\documentclass[fleqn,twoside,amsmath]{article}
\usepackage[headings]{espcrc2}

\readRCS
$Id: espcrc2.tex,v 1.2 2004/02/24 11:22:11 spepping Exp $
\usepackage{graphicx}
\usepackage[figuresright]{rotating}

\newcommand{\AmS}{{\protect\the\textfont2
  A\kern-.1667em\lower.5ex\hbox{M}\kern-.125emS}}

\usepackage{bm}
\newcommand{\f}{\frac}
\newcommand{\bfll}{\begin{flushleft}}
\newcommand{\efll}{\end{flushleft}}
\newcommand{\bt}{\begin{tabular}}
\newcommand{\et}{\end{tabular}}
\newcommand{\bce}{\begin{center}}
\newcommand{\ece}{\end{center}}
\newcommand{\ben}{\begin{enumerate}}
\newcommand{\een}{\end{enumerate}}
\newcommand{\be}{\begin{equation}}
\newcommand{\ee}{\end{equation}}

\newcommand{\ld}{\lambda}

\newcommand{\bk}{\bm{k}}

\def\s{{\sigma}}
\def\d{{\partial}}

\title{Spin Hall effect in Sr$_2$RuO$_4$ and transition metals (Nb,Ta)}
\author{T. {\sc Tanaka}, H. {\sc Kontani}, M. {\sc Naito}, D.S. {\sc Hirashima}, \address{Department of Physics, Nagoya University, Furo-cho, Nagoya 464-8602, Japan.} K. {\sc Yamada} \address{Engineering, Ritsumeikan University, 1-1-1 Noji Higashi, Kusastu, 
Shiga 525-8577, Japan.} and J. {\sc Inoue} \address{Department of Applied Physics, Nagoya University, Furo-cho, Nagoya 464-8602, Japan. }}

\runtitle{Spin Hall effect in Sr$_2$RuO$_4$ and transition metals (Nb,Ta)}
\runauthor{T. Tanaka, H. Kontani, M. Naito, D.S. Hirashima, K. Yamada and J. Inoue}

\begin{document}

\begin{abstract}
We study the intrinsic spin Hall conductivity (SHC) and the $d$-orbital Hall conductivity (OHC) in metallic $d$-electron systems based on the multiorbital tight-binding model. The obtained Hall conductivities are much larger than that in $p$-type semiconductors. The origin of these huge Hall effects is the ``effective Aharonov-Bohm phase" induced by the signs of inter-orbital hopping integrals as well as atomic spin-orbit interaction. Huge SHC and OHC due to this mecahnism is ubiquitous in multiorbital transition metals.  
\end{abstract}
\maketitle

\section{Introduction}
\label{Introduction}

Transport phenomena give us significant information on the manybody electronic states and help us to understand the electronic properties in the superconducting state. Multiorbital effect is significant in many superconductors (SC). For example, superconducting state in Sr$_2$RuO$_4$ shows a prominent orbital dependent SC \cite{Sigrist}. Multiorbital effect is also important in transport phenomena. Spin Hall effect (SHE) and anomalous Hall effect (AHE) are significant examples which arise from the multiorbital effect. 

Recent experiments declared the existence of sizable SHC in various compounds. Especially, the SHC in Pt reaches 240 $\hbar e^{-1} \Omega^{-1}$cm$^{-1}$ at room temperature, which is 10$^4$ times larger than that in semiconductors \cite{Kimura}. Now SHC in various transition metals attaracts great attention. However, simple electron gas models cannot explain this experimetal facts. In ref. \cite{Kontani-Ru}, we presented the first report on the theoretical study of the SHE in transition metals: They have shown that the anomalous velocity due to the atomic degrees of freedom gives rise to the large SHC comparable to the experimetal values. Therefore, analyses based on the multiorbital tight-binding model are indipensible to elucidate the origin of the huge SHC in transition metals. Later, refs. \cite{{Kontani-Pt},{Nagaosa}} reproduced the SHC in Pt theoretically.

In this paper, we study the intrinsic spin Hall effect (SHE) and $d$-orbital Hall effect (OHE)  based on a realistic tight-binding model. We first discuss the SHE in Sr$_2$RuO$_4$, which is a famous triplet superconductors at T$_{c}$=1.5. Next, we discuss the SHE in Nb and Ta, which are superconductors at T$_{c}$=9.23 for Nb and T$_{c}$=4.39 for Ta. 
The magnitude of obtained SHC are comparable to that in Pt. The theoretical technique developed in this study will serve to elucidate the origin of large SHC in other $d$-electron systems.

\section{SHE in Sr$_2$RuO$_4$}
\label{Sr$_2$RuO$_4$}

In this section, we study the SHE in Sr$_{2}$RuO$_{4}$, 
where the metalicity appears in two-dimensional RuO$_{2}$
planes, and the Fermi surface is composed mainly of $t_{2g}$ 
($d_{xz},d_{yz},d_{xy}$) orbitals.
The tight-binding model for Sr$_{2}$RuO$_{4}$, which we call 
the $t_{2g}$-model, is introduced in ref. \cite{Sigrist}. 

Hereafter, we denote $xz=1$, $yz=2$, $xy=3$.
Using this presentation, the matrix element of the Hamiltonian 
without spin-orbit (SO) interaction is given by \cite{Sigrist}
\begin{eqnarray}
{\hat H}_{0}=\left(
\begin{array}{ccc}
 \xi_1(k) & g(k) & 0 \\
 g(k) & \xi_2(k) & 0 \\
 0 & 0 & \xi_3(k) 
\end{array}
\right), \label{eqn:H0}
\end{eqnarray}
where the first, the second and the third row (column) 
correspond to $xz$, $yz$ and $xy$, respectively.
$\xi_1=-2t\cos k_x$, $\xi_2=-2t\cos k_y$, and 
$\xi_3=-2t_3(\cos k_x+\cos k_y)-4t_3^{\prime }\cos k_x\cos
k_y+\xi_3^{0}$ are intraorbital kinetic energies;
$t$ is the nearest neighbor $d_{xz}$-$d_{xz}$ ($d_{yz}$-$d_{yz}$) 
hopping along $x$ ($y$)-axis, 
and $t_3$, $t_3'$ are the nearest and the second nearest neighbor 
$d_{xy}$-$d_{xy}$ hoppings, respectively. 
Here, a constant $\xi^0_3$ is included in $\xi^3$ to adjust the number of electrons $n_l$ on $l$-orbital.
We note that the interorbital kinetic energy $g=-4t'\sin k_x\sin k_y$,
which breaks the mirror symmetry with respect to $k_x$- and $k_y$-axes,
causes the large anomalous velocity \cite{Kontani06}.
This is the origin of huge SHE.
Next, we consider the SO interaction $H_{SO}=\sum_{i} \ld \bm{l_{i}} \cdot \bm{s_{i}}$. Since the SO interaction mixes electrons with different spins, $H_{SO}$ is given by $6\times 6$ matrix:  
\begin{eqnarray}
{\hat H}_{\rm SO}= \f{\ld\hbar^2}{2} \left(
\begin{array}{cccccc}
 0& -i &0 & 0& 0& i \\
 +i& 0 &0 & 0& 0& -1 \\
 0& 0 &0 & -i& 1& 0 \\
 0& 0 &i & 0& i & 0 \\
 0& 0 &1 & -i& 0& 0 \\
 -i& -1 &0 & 0& 0& 0 \\
\end{array}
\right),
\end{eqnarray}
where the first three rows (columns) correspond to 
$xz\uparrow$, $yz\uparrow$ and $xy\uparrow$, 
and the second three rows (columns) correspond to 
$xz\downarrow$, $yz\downarrow$ and $xy\downarrow$, respectively.
As a result, the total Hamiltonian 
${\hat H}_{\rm tot}= {\hat H}_0+{\hat H}_{\rm SO}$
is given by $6\times6$ matrix. According to ref. \cite{Sigrist}, 
we put $t=1$, $t^{\prime }=0.1$, $t_3=0.8$, $t_3^{\prime }=0.35$, 
and assume that $t\approx 0.2$eV and ${\lambda }\sim
0.2t$.

Here, charge current operator for $\mu$-direction ($\mu=x,y$)
is given by \cite{Kontani06}
\begin{eqnarray}
& &{\hat j}_{x}^{\rm C} = -e \frac{\d {\hat H}}{\d k_x}
=-e \left(
\begin{array}{ccc}
 v_x & v_x^a & 0 \\
 v_x^a & 0 & 0 \\
 0 & 0& v_x^z \\
\end{array}
\right), \label{eqn:J}
\end{eqnarray}
where $v_x= \d \xi_1/\d k_x$ and $v_x^z= \d \xi_3 /\d k_x$.
The interorbital velocity $v_x^a= \d g/\d k_x= -4t' \sin k_y \cos k_x$
is called the ``anomalous velocity'', which is the origin of the 
Hall effects \cite{Kontani-Ru,Kontani06}.
Since $v^a_x$ has the same symmetry as $k_y$,
$\langle v_x^a v_y \rangle$ can remain finite after the $\bk$-summations.
Next, the $\s_z$-spin current and the $l_z$-orbital current are given by
${\hat j}_x^{\rm S}= \{ {\hat j}_x^{\rm C}, {\hat s}_z \}/2$ and 
${\hat j}_x^{\rm O}= \{ {\hat j}_x^{\rm C}, {\hat l}_z \}/2$, respectively \cite{Kontani-Ru}. 
In the present model, current operators are also given by $6\times 6$ matrix.

Now, we show the numerical results. We calculate the intrinsic SHC and OHC in the presence of local impurities using linear responce theory. According to the linear responce theory \cite{Sterada}, the SHC and OHC is composed of the ``Fermi surface term (I)" and ``Fermi sea term (II)". 

\begin{figure}[!htb]
\begin{center}
\includegraphics*[scale=0.35,viewport=0 0 530 430]{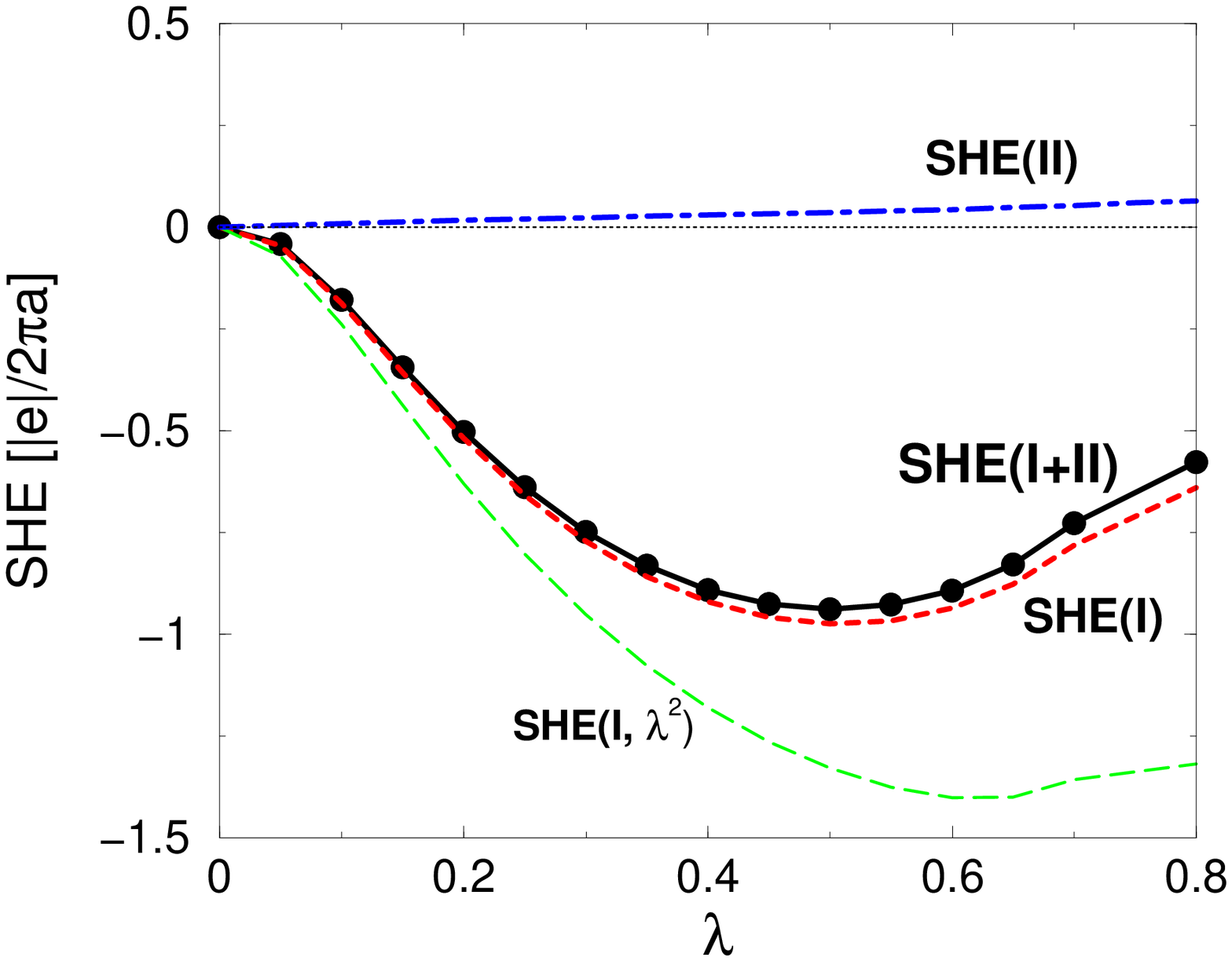}
\includegraphics*[scale=0.35,viewport=0 0 530 430]{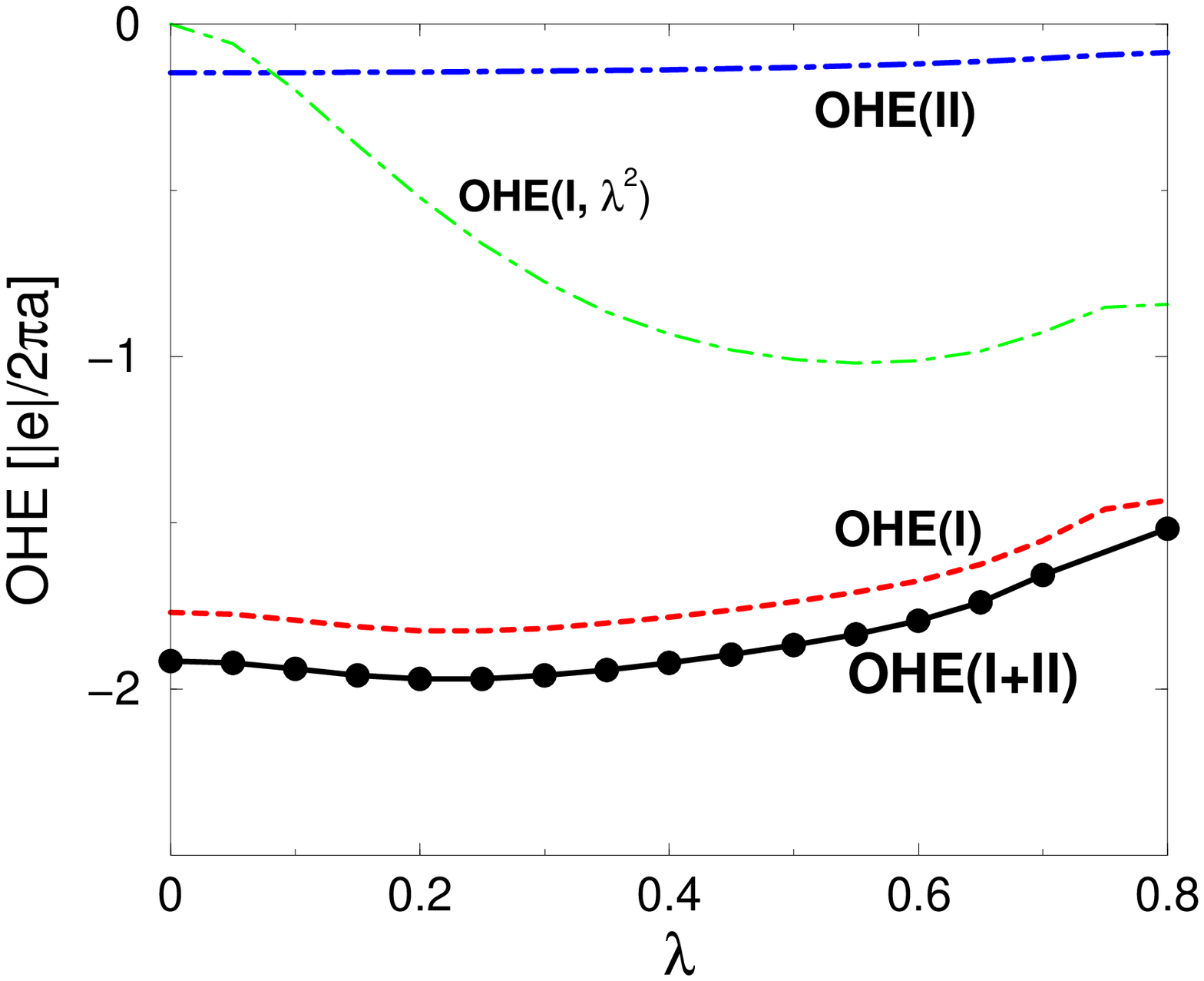}
\caption{ $\ld$-dependence of the (top) SHC and (bottom) OHC in $t_{2g}$-model for Sr$_2$RuO$_4$ [n=4]. A typical value of $\ld$ for Ru$^{4+}$-ion corresponds to 0.4.}
\label{fig:lambda}
\end{center}
\end{figure}
Figure \ref{fig:lambda} shows the $\ld$-dependence of the 
SHC and OHC for $t_{2g}$-model. Here, we set the number of electrons $n_l$ on $l$-orbital as
$n_{1}= n_{2}= n_{3}= 4/3$.
We use the Born approximation since a tiny residual resistivity 
in Sr$_2$RuO$_4$ suggests that the impurity potentials are small.
The Fermi sea terms (II) of both the SHC and OHC
are much smaller than the Fermi surface terms (I) \cite{Kontani06}.
The total SHC (OHC) is given by I+II.
Here, $1.0 \ [|e|/2\pi a]$ corresponds to 
$\approx 670 \ [\hbar/|e|]\Omega^{-1}{\rm cm}^{-1}$
if we put the interlayer distance of Sr$_2$RO$_4$; $a\approx6$\AA.
The obtained SHC and OHC for a typical values of $\ld\sim0.2$
are much larger than values in semiconductors,
because of the large Fermi surfaces and the large SO interaction
in transition metal atoms.

\begin{figure}[!htb]
\begin{center}
\includegraphics*[scale=0.4,viewport=0 0 400 300]{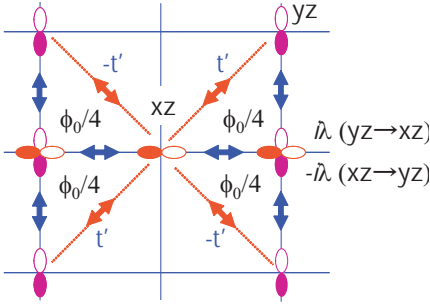}
\caption{``Effective magnetic flux" in $(xz,yz)$-model for $\downarrow$-spin electron.}
\label{fig:flux}
\end{center}
\end{figure}
In Fig. \ref{fig:flux}, we give an intuitive reason why SHC appears in $(xz,yz)$ model, which is given by dropping the third line and column in eq. (\ref{eqn:H0}). By considering the signs of interorbital hopping integrals and matrix elements of spin-orbit interaction, we can verify that a clockwise (anti-clockwise) movement of a $\downarrow$-spin electron along any triangle of half unit cell causes the factor $+i$ ($-i$). This factor can be interpreted as the Aharonov-Bohm phase factor $e^{2\pi i\phi/\phi_0}$  [$\phi_0=hc/|e|$], where $\phi$ represents the effective magnetic flux
$\phi= \oint{\bf A} d{\bf r}=\pm \phi_0/4$. Since the effective flux for $\uparrow$-spin electron is opposite in sign, electrons with different spins move to opposite direction. Therefore, the effective magnetic flux gives rise to the SHC of order $O(\lambda)$. Large SHE and OHE due to such effective flux will be realized in various multiorbital transition metal complexes.

\section{SHE in transition metals}
\label{transition}

In this section, we study the SHE in Nb and Ta. They have a body-centered cubic (bcc) structure with lattice constant $a=3.3$\AA.
To describe the electronic structure in Nb and Ta, we use the Naval Research Laboratory tight-binding (NRL-TB) model \cite{NRL} within nine orbitals; $5s,5p,4d$ for Nb and $6s,6p,5d$ orbitals for Ta. 
In the presence of SO interaction $\ld \sum_{i} \bm{l_i}\cdot \bm{s_i}$ for $d$ electrons, the total Hamiltonian is given by 
\begin{eqnarray}
\hat H =
\left(
\begin{array}{cc}
\hat H_0 + \ld \hat l_z/2 & \ld (\hat l_x - i \hat l_y)/2 \\
\ld (\hat l_x + i \hat l_y)/2 & \hat H_0 - \ld \hat l_z/2
\end{array}
\right)
\end{eqnarray}
where $\hat H_0$ is a $9\times 9$ matrix given by NRL-TB model. The matrix elements of $\bm{l}$ are given in ref. \cite{Kontani-Pt}. We set the SO coupling constant $\ld$ by use of ref. \cite{Herman}: $\ld$=0.006 Ry for 4$d$ electron in Nb, and 0.023 Ry for 5$d$electron in Ta.
We verified that the obtained band structures agree well with the results of a relativistic first-principles calculation near the Fermi level.


\begin{figure}[!htb]
\begin{center}
\includegraphics[width=0.9\linewidth]{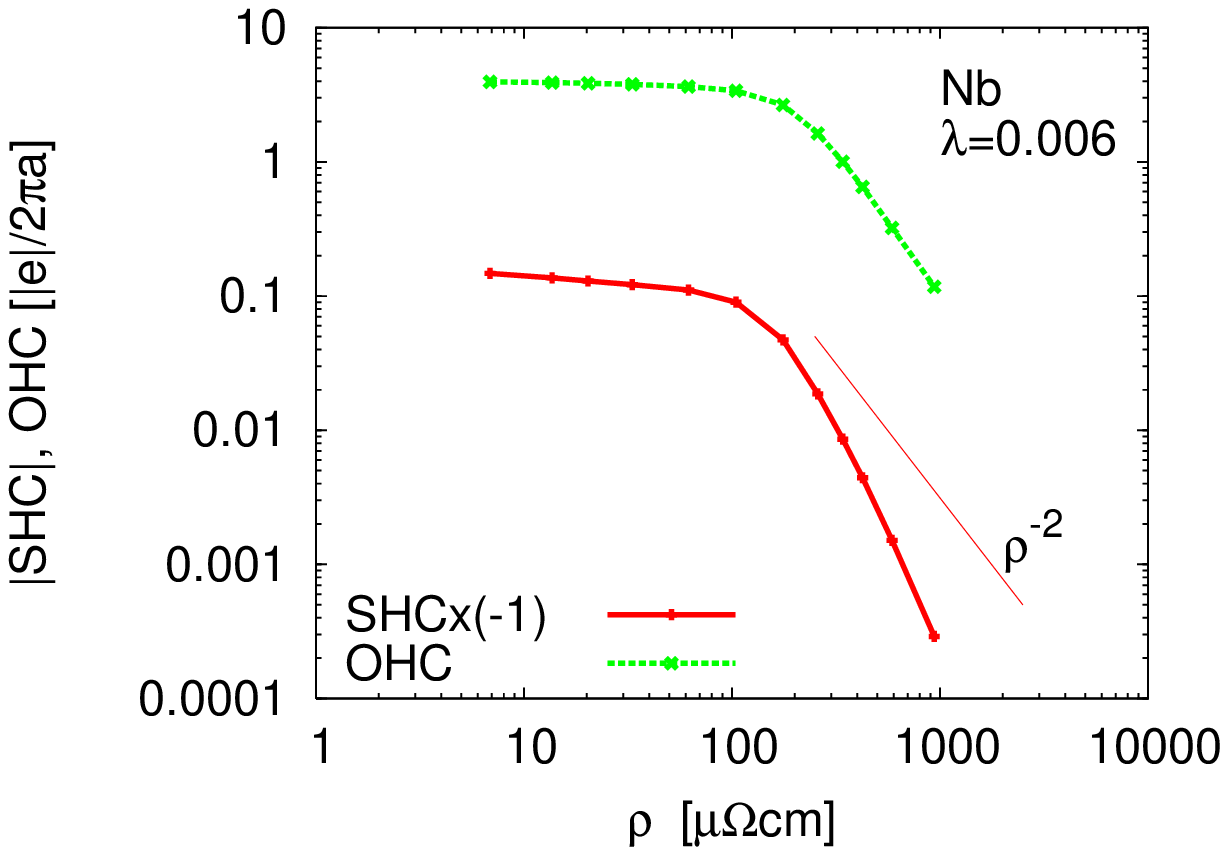}
\includegraphics[width=0.9\linewidth]{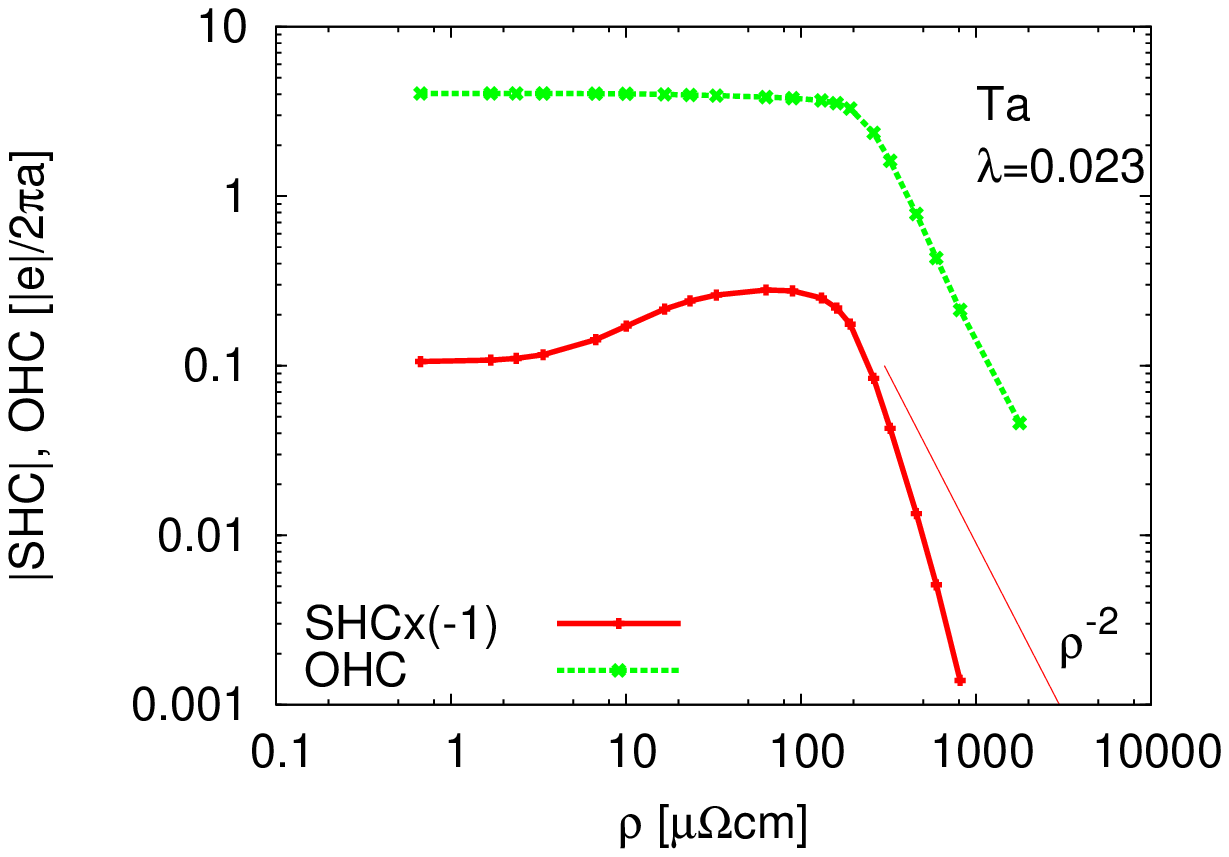}
\caption{$\rho$-dependence of SHC and OHC in Nb and Ta. Note that 1 [$|e|/2\pi a$]$\approx1000 \hbar e^{-1} \cdot \Omega^{-1}$cm$^{-1}$ if we put the lattice constant  $a=4$\AA.}
\label{fig:rho}
\end{center}
\end{figure}
Now, we perform the numerical calculations for the SHC and OHC. 
Fig. \ref{fig:rho} shows the resistivity ($\rho$) dependence of SHC and OHC in Nb and Ta. We find that the SHCs take large negative values in Nb and Ta. Note that the SHC in Pt is opposite in sign \cite{{Kontani-Pt},{Nagaosa}}.
In usual, intrisic SHC is independent of resistivity in the low resistive regime $(\rho \leq 50 \mu\Omega$cm), whereas it decreases approximately in proportional to $\rho^{-2}$ in the high resistive regime \cite{Kontani06}. We see that this coherent-incoherent crossover takes place in Nb. However, the obtained SHC in Ta decreases as $\rho$ decreases even in the low resistive regime. We find that this anomalous behavior arises when accidental degenerate points exist slightly away from the Fermi level \cite{Tanaka}. 

\section{Summary}
\label{Summary}

In summary, we studied the SHE and OHE in Sr$_2$RuO$_4$ and transition metals such as Nb and Ta. We found that huge SHE and OHE originte from the ``effective Aharonov-Bohm phase" induced by the angular momentum of the atomic orbitals.
The present study strongly suggests that ``giant SHE and OHE" are ubiquitous in multiorbital $d,f$-electron systems with atomic orbital degrees of freedom. In near future, the novel field of SHE and OHE will be extended to wide variety of materials.


\end{document}